\documentclass[epj]{svjour}

\usepackage{epsfig} 

\begin{document}

\title{External fluctuations in front dynamics with inertia:
The overdamped limit}

\author{J.\ M.\ Sancho\inst{1} and Angel S\'anchez\inst{2}}

\institute{
     \inst{1} Departament d'Estructura i Constituents de la Mat\'eria,
Universitat de Barcelona, Avenida Diagonal 647, 08028 Barcelona, Spain \\
     \inst{2} Grupo Interdisciplinar de Sistemas Complicados (GISC),
Departamento de Matem\'aticas, Universidad Carlos III de Madrid
Avenida de la Universidad, 30, 28911 Legan\'es, Madrid, Spain}

\date{Received: \today / Revised version: }

\abstract{
We study the dynamics of fronts when both inertial effects
and external fluctuations are taken into account.
Stochastic fluctuations are introduced as multiplicative noise
arising from a control parameter of the system. Contrary to
the non-inertial (overdamped) case, we find that important features
of the system, such as the velocity selection picture, are not
modified by the noise. We then compute the overdamped limit of
the underdamped dynamics in a more careful way, finding that it does not
exhibit any effect of noise either.
Our result poses the question as to whether or not external
noise sources can be measured in physical systems of this kind.  
}


\maketitle

\section{Introduction}
Front propagation is being the subject of
very active research in the last few years:
Indeed, the problem of the
selection of the front velocity is a paradigm of the dynamical 
selection mechanisms arising in a large number of physical, chemical
and biological systems with a certain kind of instability
(see \cite{ch} and references therein). 
One of the important questions into which interest has been focused is 
how the deterministic front scenario is modified by the presence of noise. 
In this context, the effect
of stochastic fluctuations on front dynamics and the modification
of its deterministic features have been considered
by several authors. A detailed summary of those results, which were
mostly devoted to the changes of the front velocity and the spreading
of the front position, can be found in \cite{book}.

The present work addresses a related problem which, on the other hand,
arises naturally from the above line of research: 
Is the influence of (external) noise on front
propagation the same if inertial effects (as far as we know, neglected
in previous work) are taken into account?  
As is well known, including inertia leads to a description
in terms of a damped, hyperbolic partial differential 
equation. It is important to note that first, a model like this arises
when one considers a more realistic jump process for the individuals 
whose probability density is described by the partial differential 
equation; and, second, that hyperbolic equations of this kind 
describe many actual physical phenomena, such as, e.g., population
dynamics, nonlinear transmission lines, cell motion, branching
random walks, dynamics of ferroelectric domains, and others
\cite{struc,do,merlin,vm,gallay,erzan,scott}.
The r\^ole of inertia in the scenario of (deterministic)
front propagation has been studied recently \cite{sancho99} (see also
\cite{vm,gallay,erzan}, and specially \cite{preprint} for a detailed study 
of the underdamped dynamics restricted to the linear regime). 
In this paper, it was proven that the different dynamical 
regimes of front propagation using deterministic parabolic models do not 
change, i.e., the values of the control parameter separating the different
regimes do not depend on the inertia parameter (``mass'').
However, it was also shown in 
\cite{sancho99} that the values of the front velocities corresponding to 
every interval of the control parameter and the spatial shape of the 
propagating fronts do depend on the inertia parameter.

As stated above, our explicit objective will now be
the study of the r\^ole of multiplicative noise and its comparison to the 
results in \cite{Joan1,Joan2}, in order to understand the interplay of inertia 
and external stochastic perturbations.  Accordingly, 
we undertake the study of a hyperbolic partial diferential 
equation with a multiplicative noise term, used to model front 
dynamics subjected to both inertia and external 
fluctuations. Opposite to the deterministic case, in which 
for very small 
mass or inertia a naive adiabatic elimination procedure 
(i.e., leaving the second time derivative term out) 
gives a parabolic equation which describes very accurately 
the front dynamics \cite{sancho99},
we will prove that this is not at all the case when noise is present. 
As a matter of fact, the 
starting hyperbolic equation {\em with noise in the Stratonovi\v c 
interpretation} transforms,  
upon adiabatic elimination, into a parabolic 
equation with an extra term coming from the noise. Furthermore, the 
so obtained equation turns out to be equivalent to the usual parabolic
equation if interpreted in the {\em It\^o sense}. 
In addition, we include numerical simulation results confirming this 
unexpected result.  
We report on these results along the 
following 
scheme: We begin by presenting our model and by briefly summarizing 
what is known about the noise influence in the overdamped case. 
Subsequently, we concern ourselves
with the study of the externally perturbed
(stochastic) case. We conclude by summarizing our main findings 
and discussing their implications. 

\section{Model definition and notations}
Let us begin by introducing the purely deterministic problem. 
Generally speaking, the
situation which we are interested in 
is generally modeled by the hyperbolic equation
\begin{equation}
\label{1}
\phi_{tt} +  \alpha \phi_{t} = D \phi_{xx} + \tau^{-1} f(\phi,a)  
\end{equation}
where $\alpha$ is the friction (dissipation), 
$D$ is the diffusion coefficient, $\tau$ 
is the characteristic time of the reaction term and $a$ is the 
external control parameter of the nonlinear reaction term $f(\phi,a)$.
The first step will be the reduction of the number of 
parameters by introducing the change of variables
$t\rightarrow \tau \alpha t$, $x\rightarrow \sqrt{\tau D }x$;
our initial model, Eq.\ (\ref{1}), reduces then to 
\begin{equation}
\epsilon \phi_{tt} + \phi_{t} = \phi_{xx} + f(\phi,a),
\label{hyperbolic}
\end{equation}
where a new parameter (the ``mass''), 
$\epsilon = ( \tau \alpha^2)^{-1} $, appears. We note 
that the information regarding both the characteristic reaction time and 
the dissipation coefficient is contained now in $\epsilon$.
With this new notation, 
the parabolic or overdamped limit is obtained by letting $\epsilon \to
0$ \cite{gallay}, which leads to (note that it is a singular limit)
\begin{equation}
\phi_{t} = \phi_{xx} + f(\phi,a).
\label{parabolic}
\end{equation}
We will refer to this procedure along the paper as {\em naive adiabatic 
elimination}.

For the sake of definiteness, we take as a representative example of
nonlinear reaction term 
\begin{equation}
f(\phi,a) = \phi(1-\phi)(a+\phi).
\label{reaction}
\end{equation}
Such a term can be obtained from a local (bistable) potential, 
$f(\phi,a) = - V'(\phi)$, with
\begin{equation}
V(\phi) = - \frac{a}{2} \phi^2 -\frac{1-a}{3} \phi^3 + \frac{1}{4} \phi^4.
\end{equation}
It is then straightforward to show that 
the steady states are, $\phi_1=0, \phi_2=1$ and $ \phi_3=-a$. We are 
interested in those solutions which are front-like (kinks) conecting the 
(unstable if $a>0$ and metastable otherwise)
state $\phi_1=0$ with the globally stable state $\phi_2=1$.
Consequently, we supplement Eqs.\ (\ref{hyperbolic}) and (\ref{parabolic})
with boundary conditions $\phi(-\infty,t)=\phi_2$, $\phi(\infty,t)=\phi_1$.

Let us now move on to the stochastic version of the problem. 
As is well known, thermal (additive) noise is {\em not} expected to be 
relevant in actual, experimental contexts, as its ratio to other terms
in the governing equations can be estimated to be $10^{-9}$ \cite{ch}. 
However, in addition to thermal noise we must expect \cite{ch} multiplicative
noise sources arising from fluctuating control parameters \cite{hl}.
Examples of this case are recent experiments on the Belousov-Zabotinsky
reaction in a light-sensitive medium \cite{showalter,irene,wang}. The
fluctuating light intensity enters as a multiplicative noise in the theoretical
modelization of this chemical reaction.
According to this, noise 
is introduced in the system described so far 
through the parameter $a$, which fluctuates according to
\begin{equation}
a\, \rightarrow a\, + \xi(x,t).
\end{equation}
The noise $\xi$ is gaussian, with zero mean and correlation given by
\begin{equation}
\langle \xi(x,t) \xi(x',t') \rangle = 2\,\sigma^2 C(x -x') \delta(t-t'),
\label{noisecorr}
\end{equation}
with
$C(x -x')$ being the spatial correlation function, normalized by imposing
$\int C(x) dx = 1$. The noisy 
parabolic case considered in \cite{Joan1,Joan2} 
corresponds to the following stochastic partial differential equation:
\begin{equation}
\phi_{t} = \phi_{xx} + f(\phi,a) + g(\phi)\xi(x,t),  
\label{paranoisy}
\end{equation}
with 
$ g(\phi) = \phi (1 - \phi)$
in case $f(\phi,a)$ is given by Eq.\ (\ref{reaction}). 
Equation (\ref{paranoisy}) with noise statistical properties 
(\ref{noisecorr}),
being well known, will be taken as our reference scenario; its main
features are summarized below. Nevertheless, before 
going into those results, it is important to note that, prior to 
any other consideration,  
it is necessary to prescribe a 
mathematical interpretation of the 
noise in Eq.\ (\ref{paranoisy}).
Based on physical and mathematical grounds we will 
follow the Stratonovi\v c interpretation (see 
\cite{stoc1,stoc2} for in-depth discussions of the interpretation 
of stochastic differential equations). 
This interpretation 
fulfills two important properties from a physical point of view: 
First, it corresponds to the white noise limit of a real (nonwhite) noise,
and second, when manipulating stochastic terms, usual rules of calculus apply.
We note that in an actual experimental situation the noise has necessarily a 
finite (non zero) characteristic time. However, this time is indeed
very small compared with any 
other characteristic time of the system, and therefore 
the assumptions of white noise and Stratonovi\v c interpretation seem 
physically well founded.

\section{Stochastic results on the overdamped model}
As already mentioned in the introduction, the overdamped limit with noise,
Eq.\ (\ref{paranoisy}), has been studied recently
\cite{book,Joan1,Joan2}. It is worth summarizing here the main points 
in order to compare with our results. In addition, we will
refer to the quantities introduced in these section later along
the text. 

In the parabolic model (\ref{paranoisy}), when $-1/2 < a < 
1/2$ (metastable and nonlinear regimes of the deterministic 
equation, see \cite{vs1,vs2,vs3}), starting from a 
sufficiently localized initial front 
evolves to the nonlinear solution, and correspondingly 
the selected velocity of the front is
\begin{equation}
v_{nl}(a) = \frac{2\,a + 1 }{\sqrt{2\,(1 - 2 \sigma^2 C(0))}}. 
\label{11}
\end{equation}
On the other hand, 
in the linear regime, $1/2 < a < 1$, the velocity is given by 
\begin{equation}
v_l = 2 \sqrt{a + \sigma^2 C(0)}.
\label{12}
\end{equation}
We note that, in the three regimes, this 
velocity increases monotonously as a function of the noise intensity;
we will come back to this result below.  

\section{Stochastic perturbation }
We now proceed to analyze the case we are interested 
in, namely the
stochastic version of the hyperbolic partial differential equation 
(\ref{hyperbolic}). For convenience,  by introducing a new field variable
$\psi$, we cast it in the form
\begin{eqnarray}
\phi_{t} &=& \psi,
\nonumber\\
\epsilon \psi_{t} &=&  -\psi + \phi_{xx} + f(\phi,a) + g(\phi) \xi(x,t). 
\label{hypernoisy}
\end{eqnarray}
At this point, it is important to note that naive adiabatic elimination 
leads again to Eq.\ (\ref{paranoisy}). 

\begin{figure}
\begin{center}
\epsfig{file=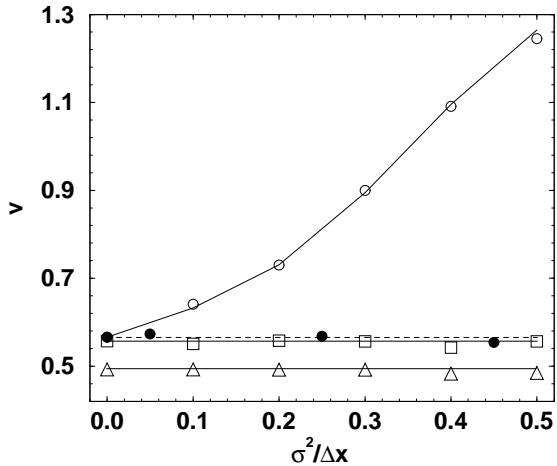, width=2.5in, angle=-90}
\caption{
Front velocity versus the effective noise intensity for 
$\epsilon=0$ [circles, parabolic case (\ref{paranoisy})], $\epsilon=0.1$ 
(squares) and $\epsilon=1$ (triangles). Filled circles correspond to the
dynamics given by Eq.\ (\ref{paranoisy2}). In all cases, $a = -0.1$;
similar results are obtained for other values of $a$. Lines 
correspond to Eqs.\ (\ref{11}), (\ref{12}), and (\ref{14}), the three
straight lower lines with $\sigma=0$; among these, 
the dashed line is the theoretical prediction 
for Eq.\ (\ref{paranoisy2}).}
\label{fig1}
\end{center}
\end{figure}

As a starting point, let us recall that
in \cite{sancho99} we proved that the velocity of the deterministic
hyperbolic model 
can be obtained from the parabolic model by using the 
transformation
\begin{equation}
v_{nl}(\epsilon,a) = \frac{1}{ \sqrt{ \epsilon + v_{nl}(a)}}.
\label{prediction}
\label{14}
\end{equation}
Therefore, our first aim here is to check whether or not
this result applies in the 
presence of multiplicative noise, i.e., whether or not we
can still use the expression (\ref{prediction}) substituting
for the deterministic velocities the stochastic ones found in
\cite{Joan1,Joan2}. 
{}From those papers, we know that the important noise effects 
come from the fact that the multiplicative noise term has {\em a non zero 
mean value}, provided the Stratonovi\v c interpretation is used. 
Within that interpretation, it can be found by 
using only usual 
stochastic calculus and a little algebra
that the mean value of the noisy term in 
Eq.\ (\ref{hypernoisy}) is \cite{book} 
\begin{equation}
\langle g(\phi) \xi(x,t) \rangle = \sigma^2 C(0) \langle \frac{\partial 
g(\phi)} {\partial \phi} \frac{ \delta \phi}{\delta \xi(x',t')} \rangle 
\Big|_{x'=x,\, t'=t} = 0.
\label{res}
\end{equation}
This result 
allows us to {\em conjecture}
that in the hyperbolic case the noise is not as relevant as 
in the parabolic case. 
In order to verify this conjecture, we carried out 
numerical simulations of Eq.\ (\ref{hypernoisy}) (with noise white in 
space, implying $C(0)=1/\Delta x$ with $\Delta x$ the spatial 
discretization step) for different choices of
the parameters, finding that the velocity is not affected by the
perturbation even for very small values of $\epsilon$.
In Fig.\ \ref{fig1} we present some 
of these numerical results. It can be clearly seen
the dramatic difference between the effects of 
noise in the two models.
It is most important to stress that the results in Fig.\ \ref{fig1} 
correspond to {\em individual realizations} of the noise; of course,
we have verified that this is the typical behavior by repeating the 
simulations very many times. This means that our conjecture that the
noise effects  and in particular those of the front
velocity are not important, is in fact true beyond mean
values, i.e., individual fronts do not change their deterministic 
velocity in the presence of noise.
For the parabolic case we see that the velocity does 
depend on the intensity of the noise, being independent in the hyperbolic case.
This is an unexpected result but, as we have just seen, it is
in complete agreement with our theoretical analysis, that leads to 
Eq.\ (\ref{res}).

Let us now discuss the reasons for this result.
It is clear from Eq.\ (\ref{res}) that 
such a null result comes from the fact that the response of the field 
$\phi$ to the noise is zero  at $t'=t$.  This is not the case in the 
parabolic case (\ref{paranoisy}), for which one finds \cite{Joan1,Joan2}
\begin{eqnarray}
\langle g(\phi) \xi(x,t) \rangle &=& \sigma^2 C(0) \langle \frac{\partial 
g(\phi)} {\partial \phi} \frac{ \delta \phi}{\delta \xi(x',t')} \rangle
\Big|_{x'=x,t'=t} = 
\nonumber\\
&=&\sigma^2 C(0) \langle \frac{\partial g(\phi)}{\partial 
\phi} g(\phi) \rangle 
\label{mean}
\end{eqnarray}
This non zero mean average is the responsable of the strong effects of 
noise in the parabolic model (\ref{paranoisy}) which were studied and 
quantified in \cite{Joan1,Joan2}. As it turns out, 
this is not the case in the 
hyperbolic model (\ref{hypernoisy}); therefore, it seems likely that
the naive 
adiabatic elimination procedure, which led to Eq.\ (\ref{paranoisy}), is 
not the proper way to take the limit $\epsilon\to 0$ in Eq.\ (\ref{hypernoisy}),
in view of the different behavior of the two models. 

In order to check this idea, we have followed an alternative,
non naive adiabatic elimination procedure \cite{sancho82} to see 
whether an equation different
{}from (\ref{paranoisy}) arises for the overdamped limit. We will 
outline here the main steps of the calculation following 
\cite{sancho82} (see \cite{stoc2} for an
alternative presentation). For the sake of simplicity,
we rewrite Eq.\ (\ref{hypernoisy}) in a more compact form:
\begin{equation}
\phi_{t} = \psi;\,\,\,\,
\psi_{t} = \frac{1}{\epsilon} \left( F(\phi) + g(\phi)\xi(x,t) \right),
\label{hypernoisy2}
\end{equation}
where $F(\phi) = f(\phi,a) + \phi_{xx}$.
Formally integrating the second expression in (\ref{hypernoisy2})
and using the first one, we 
find an integro-differential equation for the variable $\phi$:
\begin{equation}
\psi = \phi_{t} = \int_0^t dt\, \frac{ 
e^{-(t-t')/\epsilon}}{\epsilon}\left( F(\phi) + g(\phi)\xi(x,t) \right), 
\end{equation}
with the initial condiction $\psi(0)=0$.
Subsequently, formal integrations by parts are performed in order to obtain
a series 
of terms in powers of $\epsilon$, which is the situation we are 
interested in ($\epsilon$ small). {}From this formal expansion a 
Fokker-Planck equation, whose first order 
term does not depend on $\epsilon$, is obtained.
The calculation is involved, but it does not require any further physical
assumptions, it is only (lengthy) algebra. We then skip the details and
refer the reader interested in them to 
\cite{sancho82}. The final result is that the
corresponding 
Langevin equation in the Stratonovi\v c interpretation is
\begin{equation}
\phi_{t} = \phi_{xx} + f(\phi,a)  - \sigma^2 C(0) \frac{\partial 
g(\phi)}{\partial
\phi} g(\phi) +  g(\phi) \xi(x,t) 
\label{paranoisy2}
\end{equation}
We have thus obtained a different overdamped limit, for which 
one can check that, according to Eq.\ (\ref{mean}), the mean value 
of the noisy term compensates the new term in Eq.\ (\ref{paranoisy2}),
hence rendering the noise contribution null, as in the $\epsilon\neq 0$ 
case. Therefore, Eq.\ (\ref{paranoisy2}) is the physically consistent
overdamped limit of Eq.\ (\ref{hypernoisy}), insofar it exhibits the 
same behavior as this last one does for any value of the ``mass'' 
$\epsilon$.

\section{Discussion and conclusions}
In this work, we have shown analytically and numerically that 
inertial effects of any magnitude suppress 
the external noise influence on the velocity of fronts. 
Whereas the theoretical result has been conjectured by taking averages, our
numerical simulations show that the velocity of {\em individual} fronts
is unchanged by noise.
This means
that the overdamped (parabolic) equation usually employed to describe
front propagation in reaction diffusion model systems is not simply 
the limit of an underdamped (hyperbolic) version, as in that case it
is known \cite{Joan1,Joan2} that the velocity of the front does 
depend on the noise strength. In other words,  the naive prediction 
based on deterministic results, Eq.\ (\ref{prediction}), is not correct 
in the presence of multiplicative noise. We have also shown that, by 
means of a more involved adiabatic elimination procedure, it is possible
to obtain an equation for the overdamped limit which does not show 
noise effects. However, this equation differs from the usual one by an
extra term, arising from elimination, which exactly cancels the noise 
contribution to the mean velocity of the fronts. 

{}From the physical viewpoint, these results are very relevant. Indeed, 
we have seen that any amount of inertia present in the system 
will lead to front propagation at the deterministic
velocity even in the presence of external noise.
{\em 
Although the calculation
has been done for a specific model, the reasons for the vanishing of the
noise contribution are generic and do not depend on the specific choice 
of the reaction term.
}
In this context, it is then clear that, even if 
a system is
considered overdamped, generally speaking there will be some degree 
of inertia in its dynamics. In that case, the predicted changes in 
the velocity \cite{Joan1,Joan2} will not actually occur. Therefore, 
our results are in fact a criterion to establish whether a system 
(where front propagation arises)
is truly overdamped and correspondingly described by a parabolic 
equation, or in turn, it is an inertial system with very large 
damping: As we have seen, the response of the system to external noise
would be fundamentally different in both cases. This result is of 
particular importance in the study of excitable media in noisy 
environments \cite{showalter,wang}.  An additional implication of our 
findings is that, if one is interested in identifying or measuring
possible noise effects in hyperbolic problems in the class of
Eq.\ (\ref{hypernoisy}), 
the front velocity is not a good observable and it is necessary to 
resort to other indicators, such as, e.g., the fluctuations of the 
velocity. 

Finally, another remarkable fact is that the naive para\-bolic limit 
(\ref{paranoisy}) if interpreted in the It\^o sense, it is stochastically 
equivalent to the correct limit (\ref{paranoisy2}) valid for the 
Stratonovi\v c  interpretation.  This is but a further indication that
the overdamped limit of Eq.\ (\ref{hypernoisy}) is problematic and has
to be carefully performed,
as in any other instance where, for physical reasons, multiplicative 
noise has to be considered.
On the other hand, the fact that Eq.\ 
(\ref{paranoisy}) is the consistent overdamped limit in the It\^o 
interpretation has to do as well with the known result that, in 
general, perturbative expansions such as the one needed in the adiabatic
elimination procedure have many identically zero terms when carried out
in It\^o sense \cite{stoc2}. This result poses questions of a more 
mathematical character that would be interesting to address in a 
general framework for stochastic partial differential equations. 

\section*{Acknowledgements}

We thank Esteban Moro for a critical reading of the manu\-script.
Work at Barcelona was supported by DGES (Spain) through grant PB96-0241.
Work at Legan\'es was supported by DGES (Spain) through grant PB96-0119.


\begin{thebibliography}{99}

\bibitem{ch} M.\ C.\ Cross and P.\ C.\ Hohenberg,  Rev.\ Mod.\ Phys.\
{\bf 65}, 851 (1993).

\bibitem{book} J.\ Garc\'\i a-Ojalvo and J.\ M.\ Sancho, {\em Noise
 in spatially extended systems} (Springer, New York, 1999). 

\bibitem{struc} S.\ Aubry, J.\ Chem.\ Phys.\ {\bf 62},
3217 (1975); {\bf 64} 
3392 (1976); J.\ A.\ Krumhansl and J.\ R.\ Schrieffer, 
Phys.\ Rev.\ B, {\bf 11} 3535 (1975).
   
\bibitem{do} S.\ R.\ Dunbar and H.\ G.\ Othmer, in {\em Nonlinear
oscillations in biology and chemistry}
edited by S.\ Levin (Springer, Lecture
notes in Biomathematics vol.\ 66,
Berlin, 1986).

\bibitem{merlin} S.\ Fahy and R.\ Merlin, Phys.\ Rev.\ Lett.\ 
{\bf 73}, 1122 (1994).

\bibitem{vm} V.\ M\'endez and J.\ Camacho, Phys.\ Rev.\ E {\bf 55},
6476 (1997); V.\ M\'endez and A.\ Compte, 
Physica A {\bf 260}, 90 (1998).

\bibitem{gallay} Th.\ Gallay and G.\ Raugel, {\tt patt-sol/9809007} preprint
(1998);
{\tt patt-sol/9812007} preprint (1998).

\bibitem{erzan} \"O.\ Kayalar and A.\ Erzan, Phys.\ Rev.\ E {\bf 60}, 
7600 (1999). 

\bibitem{scott} A.\ Scott, {\em Nonlinear science} (Oxford University,
Oxford, 1999), and references therein.
                                          
\bibitem{sancho99} J.\ M.\ Sancho and A.\ S\'anchez, preprint (1999).

\bibitem{preprint} U.\ Ebert and  W.\ van Saarloos, preprint (1999).  

\bibitem{Joan1} J.\ Armero, J.\ M.\ Sancho, J.\ Casademunt, A.\ M.\ Lacasta,
L.\ Ram\'\i rez-Piscina and F.\ Sagu\'es, Phys.\ Rev.\ Lett.\ {\bf 76}
3045 (1996).

\bibitem{Joan2} J.\ Armero, J.\ Casademunt, L.\ Ram\'\i rez-Piscina
and J.\ M.\ Sancho, Phys.\ Rev.\ E, {\bf 58}, 5494 (1998).

\bibitem{hl} W.\ Horsthemke and R.\ Lefever, {\em Noise induced 
transitions} (Springer, Berlin, 1985). 

\bibitem{showalter} S.\ K\'ad\'ar, J.\ Wang and K.\ Showalter,
Nature {\bf 391}, 770 (1998).

\bibitem{irene} I.\ Sendi\~{n}a-Nadal, A.\ Mu\~{n}uzuri, D.\
Vives, V.\ P\'erez-Mu\~{n}uzuri, J.\ Casademunt, L.\ Ram\'{\i}rez-Piscina,
J.\ M.\ Sancho, and F.\ Sagu\'es, 
Phys.\ Rev.\ Lett.\ {\bf 80}, 5437 (1998).
  
\bibitem{wang} J.\ Wang, S.\ K\'ad\'ar, P.\ Jung and K.\ Showalter,
Phys.\ Rev.\ Lett.\ {\bf 82}, 855 (1999).
 
\bibitem{stoc1} N.\ van Kampen, {\em Stochastic processes in physics and
chemistry} (North-Holland, Amsterdam, 1982).

\bibitem{stoc2} C.\ W.\ Gardiner,
 {\em Handbook
of stochastic methods} (2nd edition, Springer, Berlin, 1985).
       
\bibitem{vs1} G.\ Dee and J.\ S.\ Langer, Phys.\ Rev.\ Lett.\ 
{\bf 50},
1583 (1983). 
See also J.\ S.\ Langer, in {\em Chance and Matter} 
edited by J.\ Souletie, J.\ Vannimenus and R.\ Stora (North-Holland, 
Amsterdam, 1987). 

\bibitem{vs2} E.\ Ben-Jacob, H.\ Brand, G.\ Dee, L.\ Kramer and 
J.\ S.\ Langer, Physica D {\bf 14}, 348 (1985).  

\bibitem{vs3} W.\ van Saarlos, Phys.\ Rev.\ Lett.\ 
{\bf 58}, 2571 (1987);
{Phys.\ Rev.\ A} {\bf 37}, 211 (1988);
{Phys.\ Rev.\ A} {\bf 39}, 6367 (1989).

\bibitem{sancho82} J.\ M.\ Sancho, M.\ San Miguel and D.\ D\"urr, J.\
Stat.\ Phys.\
{\bf 28}, 291 (1982).

\end{thebibliography}
\end{document}